\documentstyle[12pt]{article}
\newcommand{\fr}{\frac}
\newcommand{\lb}{\label}
\newcommand{\ti}{\tilde}
\newcommand{\be}{\begin{equation}}
\newcommand{\ee}{\end{equation}}
\newcommand{\beqa}{\begin{eqnarray}}
\newcommand{\al}{\alpha}

\newcommand{\del}{\partial}
\newcommand{\eeqa}{\end{eqnarray}}

\newcommand{\Pc}{{\cal P}}
\newcommand{\Rc}{{\cal R}}
\newcommand{\Qc}{{\cal Q}}
\newcommand{\Sc}{{\cal S}}

\newcommand{\omo}{\omega}
\newcommand{\oq}{[\omega ]_q}

\newcommand{\nt}{\ti{\nu}}

\begin{document}
\begin{flushright}
MRC.PH.TH--6/95 

hep-th/9510015
\end{flushright}

\vspace{1cm}

{\Large {\bf 
\noindent
q--Schr\"{o}dinger Equations for 
$V=u^2+ 1/u^2$ and  Morse Potentials 
in terms of the q--canonical Transformation } }

\vspace{1.5cm}

\noindent
\"{O}. F. DAYI$^{a,}$\footnote{E-mail address:
dayi@yunus.mam.tubitak.gov.tr.}, 
I. H. DURU$^{a,b,}$\footnote{E-mail address:
duru@yunus.mam.tubitak.gov.tr.}

\vspace{.5cm}

{\small {\it
\noindent
a) T\"{U}B\.{I}TAK-Marmara Research Centre,
Research Institute for Basic Sciences,
Department of Physics,
P.O.Box 21, 41470 Gebze, Turkey,  

\noindent
b)Trakya University, 
Mathematics Department, 
P. O. Box 126, Edirne,  Turkey. } }

\vspace{1.3cm}

{\small
\begin{center}
{\bf Abstract}
\end{center}

\vspace{.2cm}

The realizations of the Lie algebra
corresponding to the dynamical symmetry group  $SO(2,1)$ 
of the
Schr\"{o}dinger equations for the Morse and the
$V=u^2+ 1/u^2$  potentials  were known to be related
by a canonical transformation. q--deformed
analog of this transformation 
connecting    two different realizations of
the $sl_q(2)$ algebra is presented. By the virtue of the
q--canonical transformation a q--deformed
Schr\"{o}dinger equation for the Morse potential
is obtained from the q-deformed
$V=u^2+ 1/u^2$  Schr\"{o}dinger equation.
Wave functions and  eigenvalues of the 
q--Schr\"{o}dinger equations yielding
a new definition of the q--Laguerre polynomials
are studied. 

\vspace{2cm}

\pagebreak

\section{Introduction}

q--harmonic oscillators are the most extensively studied 
q--deformed systems \cite{har}. However,
there is a limited literature on the
q--deformations of  other systems. In a recent 
work a new approach was adopted
for defining new q--deformed Schr\"{o}dinger equations
in a consistent manner with the 
q--oscillators \cite{dd}: 
q--deformation of the one dimensional Kepler 
problem was obtained by the help
of q--deformed version of the 
canonical transformation connecting the   
$x^2$ and $1/x$
potentials. 

In this work we attempt to study the q--deformation of another
problem, namely the Schr\"{o}dinger equation for the Morse 
potential. Note that,
the Morse potential problem
in its undeformed form can already
be considered as a kind of q--deformed oscillator \cite{bd}.
What we are doing here
is the q--deformation of the Schr\"{o}dinger equation
for the Morse potential
itself. It is the continuation
of our program for defining 
q--deformed Schr\"{o}dinger equations consistently with the
q--oscillators: it has been known for sometime  
that the Morse potential and the one dimensional oscillator with an extra 
inverse square potential are connected by a point canonical
transformation \cite{ih83}. Taking the advantage of this fact,
we start with the q--Schr\"{o}dinger 
equation written for the potential
$V=u^2+ 1/u^2;\ u\geq 0,$
which is essentially the radial 
equation for the two dimensional oscillator in polar
coordinates. We then introduce a q--canonical transformation
relating the above potential to the Morse potential       
$V=e^{-x}-e^{-2x};\ -\infty < x < \infty ,$
which enables us to obtain
a q--Schr\"{o}dinger equation for the latter potential.

q--deformation of the Morse potential is also discussed in 
\cite{cg} in terms of ladder and shift operators, where
an application to H$_2$ molecule is given.

Obtaining the  solutions of q--Schr\"{o}dinger equations
in general is not easy. 
For q--harmonic oscillators the problem is somehow less
complicated and there can be more than one approach \cite{min}-\cite{ww}.
For example in  \cite{min} the series solution of the
undeformed problem is generalized 
in such manner that the resulting q--deformed recursion relations
can be handled. 
When a similar generalization is performed for the q--deformed
Schr\"{o}dinger equation for 
$V=u^2+ 1/u^2$ potential one obtains an equation defining
the q--Laguerre polynomials which differs from the ones
available in the literature \cite{ql}.
The difference is discussed in section 6.

In section 2 we review the 
transformation connecting the Schr\"{o}dinger 
equations of the  $V=u^2+1/u^2$ and the Morse potentials.
We also present the canonical transformation
between the realizations of the     
$sl(2)$ algebra corresponding to these potentials
which both have the same dynamical group $SO(2,1).$

In section 3 we introduce the q--deformation of 
the phase space variables suitable to 
obtain two  realizations of the  $sl_q(2)$
algebra without altering the form of the undeformed
generators except some overall and operator ordering factors. We than 
define the q--canonical transformation 
connecting these realizations.

In section 4 we first write the q--deformed 
Schr\"{o}dinger equation for the potential $V=u^2+ 1/u^2$
which  requires a  trivial generalization of the 
q--oscillator potential $V=u^2.$
We then employ the q-canonical transformation 
of Section 3 to arrive
at a Schr\"{o}dinger equation for the q--Morse potential
$V=e^{-x}-e^{-2x}.$

In section 5 we attempt to solve the 
q--Schr\"{o}dinger equations  which define
the q--Laguerre polynomials. The general scheme is described
and illustrated for the ground and the first exited levels.

\section{From the Morse Potential to $V=u^2+1/u^2$}
\renewcommand{\theequation}{2.\arabic{equation}}
\setcounter{equation}{0}

Time independent Schr\"{o}dinger equation for the Morse potential is
$(\hbar =1)$
\be
\lb{sm}
\left( -\fr{1}{2\mu}\fr{d^2}{dx^2} +Ae^{-2x}-Be^{-x} -E^{\rm M}\right) \phi (x) =0;
\ee
$x\in (-\infty ,\infty ).$ 

By making use of the variable change \cite{ih83}
\be
\lb{vcc}
x=-2\ln u,
\ee
(\ref{sm}) becomes
\be
\lb{so}
\left( -\fr{1}{2\mu}\fr{d^2}{du^2} +4Au^2 - 
\fr{4E^{\rm M} +1/8\mu}{u^2} -4B \right) \psi (u) =0;
\ee
$u\in [0,\infty ),$ with
\be
\lb{wf}
\psi (u) = \fr{1}{\sqrt{u}}\phi (-2\ln u).
\ee

(\ref{so}) is equivalent to the Schr\"{o}dinger equation
for the one dimensional 
harmonic oscillator with an extra potential barrier 
\be
\lb{vu}
V(u)=-\fr{4E^{\rm M} +1/8\mu}{u^2}
\ee
with the energy
\be
E=4B.
\ee

The normalized wave functions and the energy spectrum of the 
Schr\"{o}dinger equation (\ref{so}) are
\be
\lb{ef}
\psi_n (u)=\sqrt{\fr{2(\mu \omo )^{\nu +1}n!}{\Gamma (\nu +n+1)}}
e^{-\fr{\mu \omo}{2}u^2}u^{\nu +\fr{1}{2}} L^{(\nu )}_n(\mu \omo u^2),
\ee
and
\be
\lb{eb}
E=\omo (n+\nu +1) =4B ;\hspace{.5cm}  n=0,1,2,\dots ,
\ee
where $L^{(\nu )}_n$ are the Laguerre polynomials and $\omo ,\ \nu$
are defined as
\be
\lb{ec}
\omo =\sqrt{8A/\mu},\  \nu =-\fr{1}{2}\sqrt{2+32\mu E^{\rm M}}.
\ee

The relation (\ref{wf}) and the transformation (\ref{vcc})
give the solution of the Morse potential Schr\"{o}dinger equation
(\ref{sm}) as
\be
\phi (x) =\sqrt{\fr{2(\mu \omo )^{\fr{4B}{\omo} -n} n!}{\Gamma (4B/\omo )}}
\exp (-\fr{\mu \omo }{2}e^{-x})\exp [(-\fr{4B}{\omo}-n-1)x]
L_n^{(\fr{4B}{\omo} -n-1)} (\mu \omo e^{-x}),
\ee
where by the virtue of (\ref{eb}) $4B/ \omo$ has been replaced by
$n+\nu +1.$ To obtain the energy spectrum
we insert (\ref{ec}) into (\ref{eb}) and solve for $E^{\rm M}:$
\be
\lb{ed}
E^{\rm M}=\fr{1}{8 \mu}\left[ (\fr{4B}{\omo}-n-1)^2 -1/2 \right] .
\ee

Dynamical symmetry group of the above systems is $SO(2,1).$
For the Morse potential the
phase space realization of the related Lie algebra $sl(2)$
is given by
\beqa
M_0 & = & -2p -i ,\nonumber \\
M_+ & = & -\fr{1}{2}e^{-x} , \lb{fga} \\
M_- & = & -2p^2 e^x +\al e^x .\nonumber
\eeqa

The realization relevant to $V=u^2+1/u^2$ potential is
\beqa
L_0 & = & up_u +\fr{i}{2}, \nonumber       \\
L_+ & = & -\fr{1}{2}u^2, \lb{dgs} \\
L_- & = & -\fr{1}{2} p_u^2 +\fr{\al}{u^2}. \nonumber
\eeqa

In terms of 
the usual relations between
the coordinates and momenta 
\[
px-xp=i,\   p_uu-up_u=i,
\]
commutation relations satisfied by the above generators can be found
to be
\beqa
{[M_0,M_\pm ]}=\pm 2iM_\pm, & [M_+,M_-]=-iM_0, &  \\
& &  \nonumber \\
{[L_0,L_\pm ]}=\pm 2iL_\pm, & [L_+,L_-]=-iL_0. &   \lb{a2}
\eeqa

Eigenvalue equation for the Morse potential hamiltonian
of (\ref{sm})
\be
\left( \fr{1}{2}p^2e^x+Ae^{-x} -E^Me^x\right) e^{-x} \phi {x} 
= Be^{-x} \phi (x) ,
\ee
is equivalent to
\be
\lb{aa1}
\left(\fr{-1}{4\mu}M_- -2AM_+\right)  \xi (x)= B\xi (x), 
\ee
with the parameter $\al$ and the eigenfunction $\xi$
are identified as
\be
\lb{aq1}
\al \equiv 4\mu E^M;\  \xi (x) \equiv e^{-x}\phi (x).
\ee

Similarly $\beta u^2 +\gamma / u^2$ potential problem
can equivalently be written as the eigenvalue equation
\be
\lb{aa2}
\left( \fr{-1}{\mu}L_- - 2\beta L_+ \right)\psi =E\psi ,
\ee
with the identification 
\be
\lb{aq2}
\al \equiv -\mu\gamma .
\ee

Note that the two identifications of $\al$ given in (\ref{aq1}) 
and (\ref{aq2})     are consistent with (\ref{vu}). 
The $1/ 8\mu u^2$ term is the operator ordering contribution
resulting from the point canonical transformation
in the Schr\"{o}dinger equation. The algebraic equations (\ref{aa1})
and (\ref{aa2}) are free of the operator ordering term.

Classically,
$sl(2)$ algebra realizations
relevant to the Morse and the $V=u^2+1/u^2$ potentials 
are connected by the canonical transformation
\be
\lb{ct}
u=e^{-x/2},\    p_u=-2 e^{x/2}p.
\ee
Indeed this canonical transformation suggests  (\ref{vcc}). Note 
that the constant terms in $M_0$ and $L_0$ which result from
the operator ordering, are not the same and they drop in the 
classical limit.

\section{q--canonical transformation}
\renewcommand{\theequation}{3.\arabic{equation}}
\setcounter{equation}{0}

q--deformation of the Schr\"{o}dinger equation for 
the potential $V=u^2+1/u^2$
can be introduced similarly to the q--harmonic oscillator.
Then, by introducing the q--canonical transformation
suggested by (\ref{ct})
we can arrive at a q--deformed Morse potential Schr\"{o}dinger
equation.

We introduce q--deformation  in terms of
q--deformed  commutation relations between the phase
space variables $p,x$ and $p_u,u.$
Note that we use the same notation for the 
undeformed and  q--deformed objects. 

There are more than one definitions of q--canonical
transformations.
Most of them were introduced in terms of some
properties of the basic
q-commutators \cite{qct}. However, they are not suitable for our 
purpose.

Let us recall the definition of q--canonical transformations given
in  \cite{dd}: 
we keep the phase space realizations
of the q--deformed generators of the dynamical symmetry group
to be formally the same as the
undeformed  generators $M_i(x,p)$ and $L_i(u,p_u),$
up to operator ordering corrections and
overall factors.
We then define the transformation
$x,p\rightarrow u,p_u$ to be the q--canonical transformation
if

\noindent
$i)$ q--algebras generated by 
$M_i(x,p)$ and $L_i(u,p_u)$ are the same and,

\noindent
$ii)$ the q--commutation relations (dictated
by the first condition) between
$p$ and $x,$ and $p_u$ and $u$ are preserved.

Let the q--deformed phase space variables satisfy $(\hbar =1)$
\be
\lb{cr1}
p_u u-qup_u =i.
\ee
We then define the q--deformation of the generators
of (\ref{dgs})  as
\beqa
L_0 & = &K^{-1}( up_u +ic). \nonumber       \\
L_+ & = &-K^{-1/2} \fr{\sqrt{q}}{1+q}u^2, \lb{dgs1} \\
L_- & = &K^{-1/2}(  -\fr{\sqrt{q}}{1+q} p_u^2 +\fr{\al}{u^2}), \nonumber
\eeqa
where     the constants are
\beqa
c=\fr{1}{q(1+q)}-\fr{(q-1)(q^2+1)}{q\sqrt{q}}\al ,  \nonumber \\
K=\fr{1}{q^2\sqrt{q}} (1+q^2) \left( \sqrt{q} +
(1-q)  (1-q^2)\al \right) . \nonumber
\eeqa
The generators (\ref{dgs1}) satisfy 
\beqa
L_0L_--\fr{1}{q^2}L_-L_0 & = & -\fr{i}{q}L_- , \nonumber \\
L_0L_+-       q^2 L_+L_0 & = & iqL_+ , \lb{qal1} \\
L_+L_--\fr{1}{q^4}L_-L_+ & = &  -\fr{i}{q^2}L_0,  \nonumber
\eeqa
which is 
the $sl_q(2)$ algebra
introduced in \cite{wit}.

To obtain another realization of the $sl_q(2)$ algebra            
let us define the q--commutator of
the q--deformed variables $p$ and $x$ to be
\be
\lb{pxc}
pe^{-x/2}-qe^{-x/2}p=-\fr{i}{2}e^{-x/2}.
\ee
We introduce the q--deformation  
of the generators (\ref{fga}) as
\beqa
M_0 & = &F^{-1}( -2p +ib), \nonumber       \\
M_+ & = &-\left[ \fr{F(q+1)}{2q^2}\right]^{-1/2} 
e^{-x}, \lb{fga1} \\
M_- & = &\left[ \fr{F}{2q(q+1)}\right]^{-1/2}(  
-\fr{\sqrt{q}}{1+q} p^2e^x +\al e^x), \nonumber
\eeqa
where
\beqa
b=-\fr{q+1}{2}+\fr{2\al}{\sqrt{q}}(1-q^4),  \nonumber \\
F=1+\fr{b(1-q^2) +1}{q} . \nonumber
\eeqa
The realization given by (\ref{fga1}) satisfies 
the $sl_q(2)$ algebra
\beqa
M_0M_--\fr{1}{q^2}M_-M_0 & = & -\fr{i}{q}M_- ,\nonumber \\
M_0M_+-       q^2 M_+M_0 & = & iqM_+ , \lb{qal2} \\
M_+M_--\fr{1}{q^4}M_-M_+ & = &  -\fr{i}{q^2}M_0,  \nonumber
\eeqa
which is the same as (\ref{qal1}).

It is obvious that both the relation between
(\ref{dgs1}) and (\ref{fga1}); and the relation between
the q--commutators
(\ref{cr1}) and (\ref{pxc}) are given by the 
transformation (\ref{ct}). Since both of the
q-deformed dynamical systems satisfy the same 
$sl_q(2)$ algebra 
(\ref{qal1}), (\ref{qal2}), we conclude that
the transformation (\ref{ct}) is 
the q--canonical transformation between two systems.

\section{q--Schr\"{o}dinger equations}
\renewcommand{\theequation}{4.\arabic{equation}}
\setcounter{equation}{0}

Once the q--canonical transformation relating the q--dynamical
systems given  by (\ref{dgs1}) and (\ref{fga1})
is found, we may proceed in the opposite direction
presented in Section 2: first define q--Schr\"{o}dinger 
equation for the $V=1/u^2+u^2$ potential and then adopt
a change of variable suggested by the q--canonical transformation
(\ref{ct}) to define a q--Schr\"{o}dinger equation for the 
Morse potential. To this end
we define the q-deformed Schr\"{o}dinger 
equation  
\be
\lb{sc1}
(-\fr{1}{2\mu} D_q^2(u) +A_q u^2 
+\fr{\al_q}{u^2}   -E_q ) \psi_q(u) =0,
\ee
by using the q--hamiltonian
\be
H_q=\fr{-1}{\mu}L_- - \mu A_q L_+,
\ee
where we used the realization $p_u=iD_q(u)$ and
\be
\lb{all}
\al_q \equiv -\fr{q^{-1/2} + q^{1/2}}{2\mu } \al .
\ee

To reproduce the undeformed equation (\ref{so}) and the 
definitions of (\ref{ec}) in the $q=1$ limit,
the constants $A_q$ and $\al_q$ should be chosen to satisfy
\beqa
A_{q=1}=\fr{1}{2}\omo^2\mu ,\nonumber \\
\al_{q=1}=\fr{-1}{2\mu} (\nu^2-1/4 ). \nonumber
\eeqa
$D_q(u)$ is the q-deformed derivative  defined as \cite{vil}
\be
\lb{qd}
D_q(u)f(u) \equiv \fr{f(u)-f(qu)}{(1-q)u} .
\ee

The variable change suggested by (\ref{ct})
\be
u=\exp (-x/2),
\ee
leads to  the q-deformed Schr\"{o}dinger equation 
for the Morse potential
\be
\lb{qsm}
(-\fr{1}{2\mu} {\cal D}_q^2(x) +
A_q  e^{-2x}
-E_qe^{-x} +\al_q)\phi_q(x) =0.
\ee
Here the wave function $\phi_q$ is given by
\be
\lb{pqx}
\phi_q(x)=e^x\psi_q(e^{-x/2}),
\ee
and the coefficient $\al_q$ is identified with the energy
of the q--Schr\"{o}dinger equation for the Morse potential
\be
\lb{qmpe}
E_q^{\rm M} \equiv \al_q.
\ee

The kinetic term is defined as
\be
\lb{cd2}
{\cal D}_q^2(x) \equiv  \fr{1}{(1-q)^2}
[1-\fr{1+q}{q}e^{2\ln q} e^{-2\ln q\del_x } +\fr{1}{q}
e^{4\ln q}e^{-4\ln q\del_x}],
\ee
which in the $q=1$ limit becomes
\[
\lim_{q\rightarrow 1}{\cal D}_q^2(x) =\fr{\del^2}{\del x^2} .
\]
Obviously, the kinetic term (\ref{cd2}) is an unusual one. The origin of
this fact lies in the observation that the undeformed 
Morse potential Schr\"{o}dinger equation 
itself can be viewed as a deformed object whose deformation
parameter is the scale of $x$ \cite{bd}.

Similar to the nondeformed case, 
by using the q--hamiltonian
\be
H_q^M \equiv   \fr{-1}{4\mu } M_- -A_q M_+
\ee
where the realization $p=i{\cal D}$ and the identification
of $\al$ as given in (\ref{all})
the q--Schr\"{o}dinger equation
(\ref{qsm}) can be written as an algebraic eigenvalue equation
\be
H\psi_q  =B \psi_q .
\ee

\section{Solutions of the q-Schr\"{o}dinger equations}
\renewcommand{\theequation}{5.\arabic{equation}}
\setcounter{equation}{0}

Inspired by the form of the undeformed solution
(\ref{ef}), we try to build the solutions of 
(\ref{sc1}) on the ground state of the 
q--Schr\"{o}dinger 
equation of
the harmonic oscillator
(obtained by setting $\al_q=0$ in (\ref{sc1})) given
in terms of the q--exponential \cite{dd}
\be
e_q(z^2) =1+\sum_{n=1}^\infty \left(
\prod_{k=1}^n 
\fr{2(1-q)}{1-q^{2k}} \right) z^{2n},
\ee
which is defined to satisfy
\[
D_q(z)e_q(z^2) =2ze_q(z^2).
\]
In terms of the constants
$\mu ,\ \omo$ and $\nt$ which can be identified, respectively, as the
mass, frequency of the harmonic oscillator and the generalization 
of $\nu ,$ let us choose
\beqa
A_q  &  =     & \fr{\mu}{2}q^{2\nt +2}\omo^{\prime 2} , \lb{aaq} \\
\al_q &  =   & \fr{1}{2\mu} [\nt +1/2]_q [\nt -1/2]_q ,\lb{aoq}
\eeqa
where 
\be
 \omo^\prime  = \fr{q^{-\nt+1/2} [\omo ]_q [2\nt +2]_q}{[\nt +3/2]_qq +
 [\nt+1/2]_q} ,
\ee
and
\be
\lb{co}
[ O  ]_q \equiv 
\fr{1-q^O}{1-q}.
\ee

One can easily verify that the wave 
function  
\be
\lb{gs}
\psi_{0,q}(u)=u^{\nt +1/2}e_q(-\mu \omo^\prime u^2/2),
\ee
is the solution of (\ref{sc1})
with the energy 
\be
\lb{gse}
E_{0,q}    =  \fr{[\omo ]_q}{2} [2\nt +2]_q .
\ee
Since (\ref{gs}) and (\ref{gse}) are 
the generalizations of the ground
state wave function 
and energy of the undeformed Schr\"{o}dinger equation,
we identify $\psi_{0,q}(u)$ as  the ground state
of the q--Schr\"{o}dinger equation (\ref{sc1}).

We then start to build the other solutions on the ground state (\ref{gs}).
Introduce the {\it Ansatz} for the n{\it th} state
\be
\lb{hsd}
\psi_{n,q} \equiv L_{n,q} (u) \psi_0(u,q^n),
\ee
where
\be
\psi_0(u,q^n) \equiv u^{\nt +1/2}e_q(\fr{-\mu \omo^\prime u^2}{2 q^{n}}).
\ee
Substituting (\ref{hsd}) into (\ref{sc1}) we obtain
\beqa
-\fr{1}{2\mu} D^2_q(u) L_{n,q} (u) +
[2]_qq^{-1}\left(\fr{\omo^\prime}{2}q^{\nt -n+1/2} u -
\fr{1}{2\mu}[\nt +1/2]_q u^{-1} \right) D_q(u) L_{n,q} (qu)   \nonumber \\
+A_qu^2\left( L_{n,q} (u) -q^{-2n}L_{n,q} (q^2u) \right)+
\fr{\al_q}{u^2}\left(L_{n,q} (u)-L_{n,q} (q^2u)  \right)   \nonumber \\
+q^{-n}E_{0,q}L_{n,q} (q^2u) -E_{n,q}L_{n,q} (u)  =0. \hspace{.4cm} \lb{dql}
\eeqa

Since we look for solutions possessing the correct $q=1$ limit,
we only deal  with 
$n={\rm even }$ eigenfunctions
and write
\be
L_{n,q}(u)=\sum_{k=0}^{n/2} a_{2k}^{(n)}u^{2k}.
\ee
Inserting the above polynomial into 
(\ref{dql}) we arrive at the three term recursion relations
\be
\lb{rr}
\Pc_{2k}^{(n)} a_{2k}^{(n)} +\Rc_{2k}^{(n)} a_{2k-2}^{(n)} 
+ \Qc_{2k}^{(n)}a_{2k-4}^{(n)}=0;\  1\leq k\leq n/2 + 1,  
\ee
where
\be
\Qc^{(n)}_2 = \Pc^{(n)}_{n+2}= 0,
\ee
and 
\beqa
\Pc_{2k}^{(n)} & = &  -\fr{1}{2\mu}[2k]_q[2k-1]_q -\fr{[2]_q}{2\mu q}
[\nt +1/2 ]_q[2k]_qq^{2k} +\al_q(1-q^{4k}), \\
\Rc_{2k}^{(n)} & = &  \fr{\omo^\prime}{2}[2]_q q^{-(\nt +n-2k +3/2)} [2k-2]_q
+q^{-n+4k-4}E_{0,q} -E_{n,q}, \\
\Qc_{2k}^{(n)} & = & A_q (1-q^{-2n+4k-8}).
\eeqa

(\ref{rr}) can be transformed into the two term
recursion relations
\be
\lb{ss}
\Sc_{2l}^1(n) a_{2l}^{(n)} +\Sc_{2l}^2(n) 
a^{(n)}_{2l-2}=0;\  1\leq l\leq n/2 + 1,  
\ee
and an $(n/2 +1)${\it th} order equation for the energy eigenvalue
$E_{n,q}$
\be
\lb{sek}
\sum_{k=0}^{n/2 +1}f_k(n,q) E_{n,q}^k=0.
\ee
Here $\Sc^{1,2}_{2l}(n),$ the coefficients
$f_k(n,q)$  and then the energy $E_{n,q}$
should be found for each $n.$
To be sure that the energy eigenvalues $E_{n,q}$
lead to the correct $q=1$ limit, we define
\be
\lb{enq}
E_{n,q}=\fr{\oq}{2} 
[2n +2\nt +2]_q +K_{n,q} ,
\ee
where $K_{n,q}$ will be found as the solution of (\ref{sek})
subject to the condition 
\[
K_{n,q=1}=0.
\]

To have an insight, let us deal with  $n=2$ case:

Solution of the q--Schr\"{o}dinger equation (\ref{sc1})
is
\be
\lb{n2s}
\psi_{2,q}=(a_0^{(2)}+a_2^{(2)}u^2 )
u^{\nt +1/2}e_q(-\fr{\mu \omo^\prime }{2 q}u^2),
\ee
whose coefficients satisfy
\beqa
\{ [2]_q +q[2]^2_q [\nt+1/2]_q + 2\mu \al_q(q^4-1) 
\} a_2^{(2)}   
+ \{ \mu [\omo ]_q q^{2\nt} [4]_q +2\mu K_{2,q} \} a_0^{(2)}   
 =   0, \hspace{.4cm} \lb{21} \\
\{ \omo^\prime [2]^2_q q^{-\nt +1/2} -\oq q^{-2}[4]_q -2K_{2,q} \} a_2^{(2)}         
+2A_q(1-q^{-4})a_0^{(2)}
 =   0. \hspace{.4cm} \lb{22} 
\eeqa
We can choose $a_2^{(2)}=1,$ then (\ref{21}) leads to
\be
a_0^{(2)} =\fr{
\{ [2]_q +q[2]^2_q [\nt+1/2]_q + 2\mu \al_q(q^4-1)\} }{
\mu [\omo ]_q q^{2\nt} [4]_q +2\mu K_{2,q} } ,
\ee
where due to (\ref{22}) $K_{2,q}$ satisfies
the second order equation
\be
\lb{efk}
K_{2,q}^2+\al (q)K_{2,q}+\beta (q)=0,
\ee
with the coefficients given by
\beqa
\al (q)=\fr{[\omo ]_q}{2}\{ \fr{[2]_q q^2 [-2\nt -1]_q}{[\nt +3/2]_q
[\nt +1/2]_q} + [4]_q(q^{-2}+q^{2 \nt})\}, \nonumber \\
\beta (q)=\fr{-1}{4\mu}A_q(1- q^{-4}
\{ [2]_q +q[2]^2_q [\nt+1/2]_q + 2\mu \al_q(q^4-1)\} ) .
\eeqa
Only one of the two solutions of (\ref{efk})  satisfy
$K_{2,q=1}=0$ condition:
\be
K_{2,q}=\fr{-1}{2}\al (q) +\fr{1}{2}\sqrt{\al^2(q)-4\beta (q)}.
\ee

The calculations for the higher states can be carried
out in a straightforward manner in spite of their
messy character.

To obtain the  energy eigenvalues of the q-Morse
problem one should solve $\nt$ in terms of $E_{n,q}$ from
(\ref{enq})  as
$\nt_{n,q} =\nt (n,\omo ,q, E_{n,q}),$ with
$E_{n,q}$ now playing the role of the coefficient of the q--Morse
potential. Inserting this into (\ref{aoq}) and using the identification
(\ref{qmpe}) one arrives at
\[
E_{n,q}^{{\rm M}}=
\fr{1}{2\mu} [\nt (n,\omo ,q, E_{n,q}) +1/2]_q 
[\nt (n,\omo ,q, E_{n,q})-1/2]_q .
\]
Obviously, the wave functions which are the solutions of 
the q--Schr\"{o}dinger equation for the Morse potential
(\ref{qsm}) corresponding to the above energy eigenvalues are
given by
\be
\lb{pqxn}
\phi_{n,q}(x)=e^x\psi_{n,q}(e^{-x/2}).
\ee

As an illustration it is enough to present the formulas for 
the ground state: 

$\nt_{0,q}$ can be solved as
\be
\nt (0,\omo ,q, E_{0,q})
=\fr{\ln \{ 1-\fr{2(1-q)}{[\omo ]_q}E_{0,q}\} }{2\ln q} -1.
\ee
Hence the q--Schr\"{o}dinger equation
(\ref{qsm})  possesses the ground state solution
\be
\lb{pqx0}
\phi_{0,q}(x)=e^x \exp\left( \fr{-x}{4\ln q}
\ln \{ 1-\fr{2(1-q)}{[\omo ]_q}E_{0,q}\} 
 +\fr{x}{2} +\fr{1}{4} \right) e_q\left(
-\fr{\mu\omo^\prime}{2} e^{-x} \right)
\ee
with the energy 
\be
E_{0,q}^{\rm M}=\fr{1}{2\mu}
\left[
\fr{\ln \{ 1-\fr{2(1-q)}{[\omo ]_q}E_{0,q}\} }{2\ln q} -\fr{1}{2}
\right]_q
\left[
\fr{\ln \{ 1-\fr{2(1-q)}{[\omo ]_q}E_{0,q}\} }{2\ln q} -\fr{3}{2}
\right]_q .
\ee

\section{Discussions}
\renewcommand{\theequation}{6.\arabic{equation}}
\setcounter{equation}{0}

Application of a q--canonical transformation enabled us to obtain 
q--Morse potential consistent with the deformed oscillator like potential
$V=u^2 +1/u^2.$ Polynomial solutions of the q--Schr\"{o}dinger
equation of the latter (\ref{sc1}) lead to a new definition of 
q--Laguerre polynomials (\ref{dql}). (\ref{sc1}) 
is an eigenvalue equation which does not involve the first q--derivatives. 
The other definitions of 
q--Laguerre polynomials \cite{ql} can be shown to lead
after a suitable coordinate change  and 
wave function {\it Ansatz} to
equations which are essentially of the form
\be
\lb{le}
D^2_q(u) \phi (u) + \nu (u) \phi (qu) -c_q \phi (u) =0,
\ee
where $c_q$ is a constant which can be vanishing. 

Obviously  (\ref{le}) leads to the Schr\"{o}dinger equation in $q=1$
limit, but it is not an eigenvalue equation as in (\ref{sc1}):
the scaling operator is equivalent to
the first order q--derivative.

Of course, one can always add some terms which are vanishing in $q=1$
limit to the q--deformed objects without altering the non-deformed 
limit. Thus, if we permit the appearance of scaling operator in 
q--Schr\"{o}dinger equation there can be infintely many varieties.
In our approach however there is only one possible definition 
of q--Schr\"{o}dinger equation.

Orthogonality properties of the q--Laguerre polynomials hence the
inner product of the related Hilbert space should be studied to
discuss the hermiticity properties of
the q--deformed objects.

\pagebreak

\newcommand{\bi}{\bibitem}

\end{document}